\documentclass[aps,prl,reprint,preprintnumbers,amsmath,amssymb,floatfix]{revtex4-1}

\usepackage{longtable}
\usepackage{morefloats}
\usepackage[dvips]{graphicx}
\usepackage{subfigure}
\usepackage{color}
\usepackage{epsfig,graphicx,amsfonts,amsbsy}
\usepackage{amsmath,amsfonts,amsthm,amssymb}
\usepackage{appendix}
\usepackage{bbm}
\usepackage{makeidx}
\usepackage{url}
\usepackage{verbatim}
\usepackage[bookmarksnumbered,pdfpagelabels=true,plainpages=false,colorlinks=true,linkcolor=blue,citecolor=blue,urlcolor=blue]{hyperref}
\usepackage[rightcaption]{sidecap}
\usepackage{array}
\usepackage{booktabs}
\usepackage{multirow}
\usepackage{bbm}
\usepackage{tabularx}
\usepackage{cancel,soul,ulem}

\newcommand{\bs}[1]{\boldsymbol{#1}}

\graphicspath{{figures/}}

\makeatletter

\begin{document}

\title{Magnonic Thermal Machines}

\author{Nicolas Vidal-Silva$^{1}$} 
\email{nicolas.vidal@ufrontera.cl}
\author{Francisco J. Peña$^{2}$}
\author{Roberto E. Troncoso$^{3}$}
\author{Patricio Vargas$^{4}$}

\affiliation{$^{1}$Departamento de Ciencias F\'isicas, Universidad de La Frontera, Casilla 54-D, Temuco, Chile}
\affiliation{$^{2}$Departamento de Física, Universidad Técnica Federico Santa María, 2390123 Valparaíso, Chile}
\affiliation{$^{3}$School of Engineering and Sciences, Universidad Adolfo Ib\'a\~nez, Santiago, Chile}
\affiliation{$^{4}$Departamento de Física, CEDENNA, Universidad Técnica Federico Santa María, 2390123 Valparaíso, Chile}

\date{\today}

\begin{abstract}
We propose a magnon-based thermal machine in two-dimensional (2D) magnetic insulators. The thermodynamical cycles are engineered by exposing a magnon spin system to thermal baths at different temperatures and tuning the Dzyaloshinskii-Moriya (DM) interaction. We find for the Otto cycle that a thermal gas of magnons converts a fraction of heat into energy in the form of work, where the efficiency is maximized for specific values of DM, reaching the corresponding Carnot efficiency. We witness a positive to negative net work transition during the cycle that marks the onset of a refrigerator-like behavior. The work produced by the magnonic heat engine enhances the magnon chemical potential. The last enables a spin accumulation that might result in the pumping of spin currents at the interfaces of metal-magnet heterostructures. Our work opens new possibilities for the efficient leverage of conventional two-dimensional magnets.
\end{abstract}

\maketitle
{\it Introduction}.- Heat engines have been at the core of science and engineering developments since the nineteenth century \cite{bejan2017evolution,andresen1984thermodynamics,caton2012thermodynamic}. At a small scale, thermal machines perform thermodynamic cycles employing quantum systems as the working medium \cite{myers2022quantum,kosloff2013quantum,vinjanampathy2016quantum,alicki2018introduction,goswami2013thermodynamics,niedenzu2018quantum}. A remarkable example is the quantum Otto cycle composed of two isochoric and two quantum-adiabatic trajectories \cite{kosloff2017quantum,pena2020otto,thomas2011coupled}. At each stage, the system only exchanges one form of energy, heat, or work, making it an ideal platform for theoretical and practical studies \cite{rossnagel2016single,zanin2019experimental,pekola2015towards,binder2018thermodynamics,horodecki2013fundamental,narasimhachar2015low,faist2019thermodynamic,manzano2018optimal,scully2003extracting}. This cycle has been examined in a wide range of working mediums such as three-level \cite{pena2020otto} and graphene-based systems \cite{pena2020quasistatic,singh2021magic,pena2015magnetostrain,myers2023multilayer}, harmonic oscillators \cite{kosloff2017quantum}, quantum dots \cite{pena2019magnetic,de2021two}, spin systems \cite{altintas2015general,thomas2011coupled}, among others \cite{barrios2017role,altintas2015rabi,azimi2014quantum,chotorlishvili2016superadiabatic}.

In spintronics, a discipline that exploits the spin of electrons and magnets, the role of thermal properties has been limited to setting the ground for spin-angular momentum transport \cite{vzutic2004spintronics}. In magnetic insulators, the transport of spin is carried by magnons, the quanta of spin fluctuations of the order parameter, and is coupled to heat flows due to the inherent spin-lattice coupling \cite{jungfleisch2022two,cornelissen2016magnon,chumak2012direct,troncoso2020spin,cornelissen2015long,streib2019magnon}. Thus, when the local thermodynamical equilibrium of a magnon gas, defined by temperature or chemical potential, is subjected to external driving, it provides routes to detect magnonic effects through the thermal and spin conductivity in linear response theory \cite{vandaele2017thermal,cornelissen2016magnon,cornelissen2016temperature,cornelissen2018spin,olsson2020pure,wei2022giant}. For instance, under temperature gradients, thermal transport measurements yield signatures of topological magnon states \cite{wang2021topological,wang2018topological,mcclarty2022topological}, entanglement \cite{mousolou2021magnon,yuan2020enhancement}, long-distance transport \cite{cornelissen2015long,cornelissen2016magnon}, and thermal diffusion in the form of spin Seebeck effect \cite{xiao2010theory,rezende2014magnon,adachi2013theory}.
Despite this, fundamental thermodynamical behaviors, such as entropy production, adiabatic processes, caloric phenomena, or thermal cycles based on slowly varying external fields, lack deeper comprehension in magnonic systems compared to their electronic counterparts \cite{munoz2012quantum,ramos2017spin,bauer2012spin}. Specifically, magnon-based thermodynamic cycles are an unexplored issue so far that constitutes a promising arena that will boost slow-magnonics. 
\begin{figure}[h]
\centering
\includegraphics[width=8.cm]{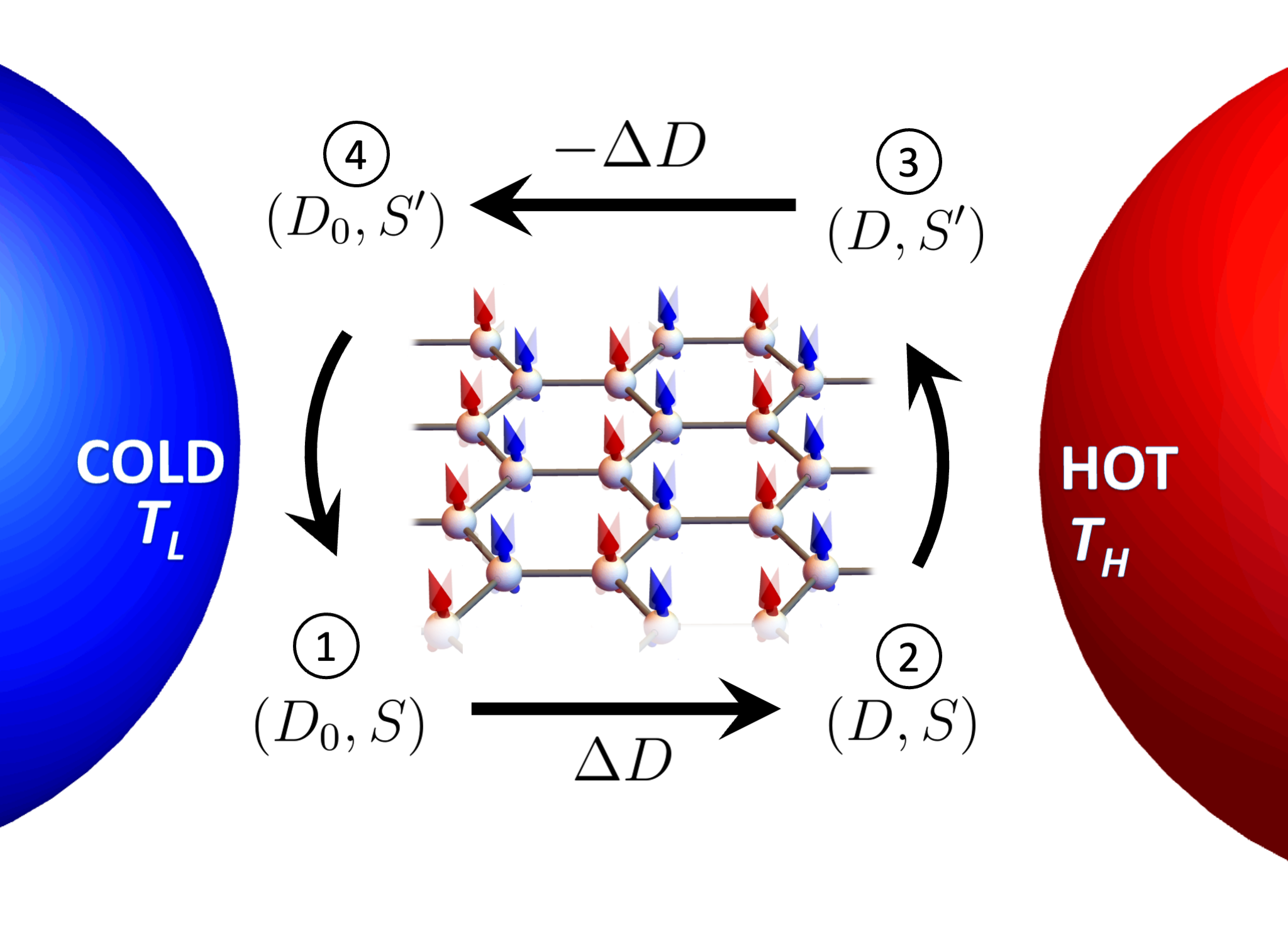}
\caption{Schematic of the magnonic quantum Otto cycle based on a magnetic honeycomb lattice with fluctuating spins. The cycle operates between hot ($T_H$) and cold ($T_C$) temperatures. In the horizontal processes, the entropy $S$ ($S'$) holds fixed, while the DM parameter $D$ ($D_0$) holds unchanged in the vertical processes.}
\label{fig:OttoCycle}
\end{figure}

In this Letter, we propose a spin-based thermal machine working with a magnonic substance that is controlled by varying the temperature and the DM parameter. The latter can be achieved by means of the application of electric fields \cite{dai2023electric,srivastava2018large,koyama2018electric} or strain \cite{gusev2020manipulation,udalov2020strain}. This thermal machine operates at a scale where thermal and spin fluctuations are relevant. The Otto cycle is proposed as an example of a thermal machine where the efficiency is maximized for specific DM coupling. Interestingly, the heat flow is inverted for a certain combination of temperature differences and DM parameter, and the system becomes a refrigerator. Considering a simple model for a two-dimensional magnon gas, the work produced by the engine can be interpreted as a spin accumulation that, in turn, might be traduced in the generation of spin currents.
\begin{figure}[thb]
\centering
\includegraphics[width=\columnwidth]{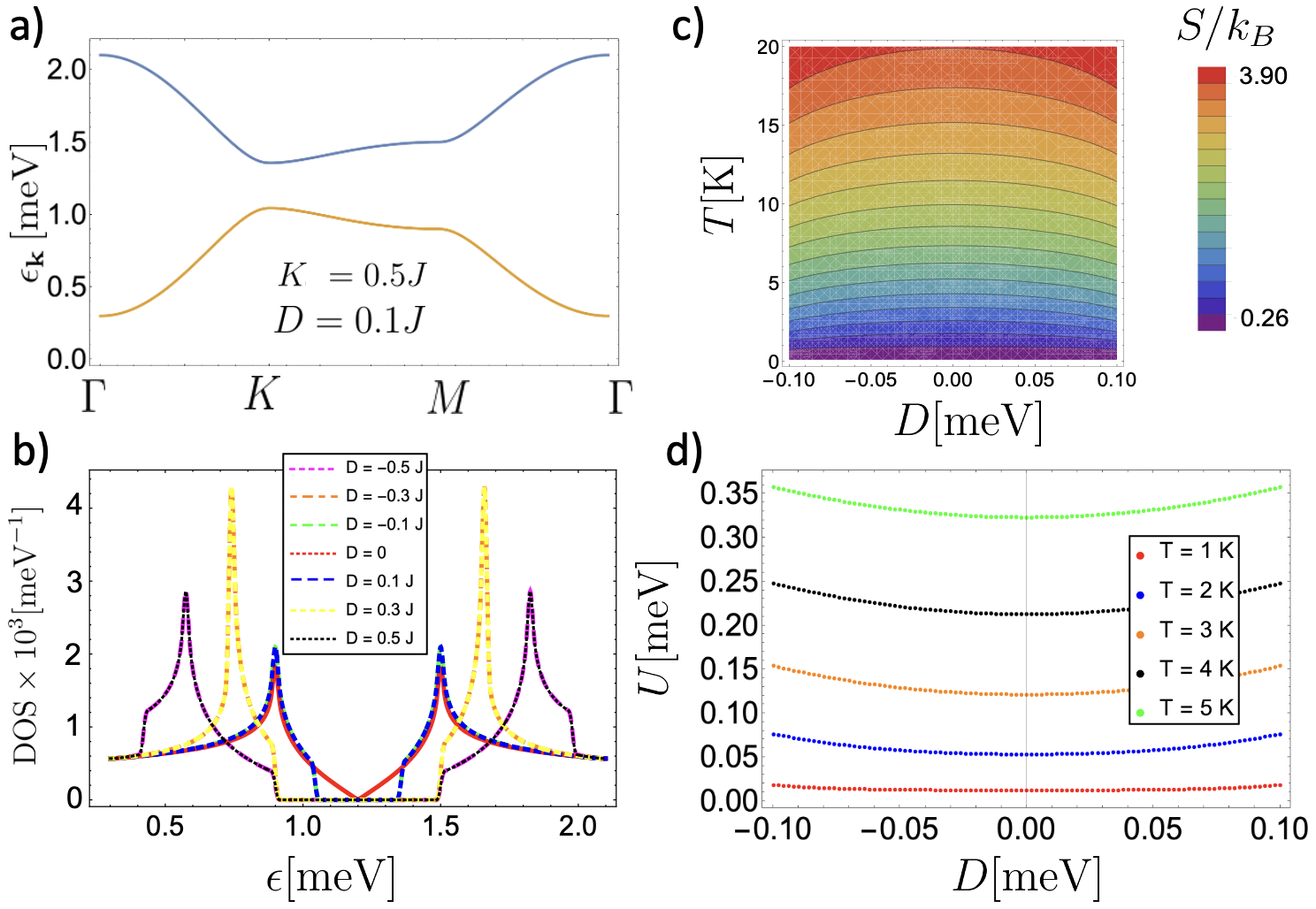}
\caption{(a) Magnon dispersion relation for $D=0.1J$ and $K=0.5J$. (b) Density of states for different values of DM coupling for $K=0.5J$. (c) Constant entropy contours $S(T,D)$ and (d) internal energy as a function of the DM parameter for different temperatures.}
\label{fig: energy-DOS-entropy}
\end{figure}

{\it Model.--} We start out by considering a two-dimensional ferromagnetic insulator. The spin system is defined on a hexagonal lattice and described by the spin Hamiltonian ${\cal H}_S=-J\sum_{\langle ij\rangle}{\bs s}_i\cdot{\bs s}_j+\sum_{\langle \langle ij\rangle\rangle}{\bs D }_{ij}\cdot\left({\bs s}_i\times{\bs s}_j\right)-\sum_{i}\left[K(s_i^z)^2+Bs_i^z\right]$, with $J$ the coupling of the nearest neighbors exchange interaction. The Dzyaloshinskii-Moriya interaction couples next-nearest neighbors spin with strength ${\bs D }_{ij}=D{\nu}_{ij}{\bs e}_{z}$, being $\nu_{ij}=\pm 1$, $K$ the easy-axis magnetic anisotropy \cite{hidalgo2020magnon,wang2018anomalous,kim2022topological}, and the external magnetic field $B$ along the $z-$direction. We focus on small spin deviations about the ground state within a linear spin-wave theory. Using the Holstein-Primakoff (HP) mapping \cite{holstein1940field}, spin fluctuations are represented by bosonic excitations via ${\bs s}^{+}_{i}=(2s-a^{\dagger}_{i}a_{i})^{1/2}a_{i}$, ${\bs s}^{-}_{i}=a^{\dagger}_{i}(2s-a^{\dagger}_{i}a_{i})^{1/2}$ and ${\bs s}^{z}_{i}=s-a^{\dagger}_{i}a_{i}$, where $a_{i} (a_{i}^{\dagger})$ is an operator that annihilates (creates) a magnon state at site $i$. In momentum space, the Hamiltonian for non-interacting magnons is ${\cal H}_m = \sum_{\bs k}\Psi_{\bs k}^{\dagger}{\cal H}_{\bs k}\Psi_{\bs k}$, where the field operator is $\Psi_{\bs k} = (\alpha_{\bs k},\beta_{\bs k})^T$, with $\alpha$ and $\beta$ acting in the sub-lattices $\mathcal{A}$ and $\mathcal{B}$, respectively. In addition, ${\cal H}_{\bs k}=\left[\Omega \mathbb{I}+{\bs h_k}\cdot\bs{\tau}\right]$ where $\Omega = 3Js + 2Ks +B$, $\bs{\tau}$ is the vector of Pauli matrices, and the vector field ${\bs h}_{\bs k}=sJ\sum_{i}\left(-\cos\left[{\bs k}\cdot{{\bs\delta}_{i}}\right],\sin\left[{\bs k}\cdot{{\bs \delta}_{i}}\right],2D\sin\left[\bs{k}\cdot\bs{\delta}_{i}^{n}\right]/J\right)^T$,
with $\bs{\delta}_{\eta}$ and $\bs{\delta}_{\eta}^{n}$ the nearest and next-nearest neighbors, respectively. The two-band bulk magnon spectrum are given by $\epsilon^{\pm}({\bs k})=\Omega\pm\sqrt{\bs{h}_{\bs k}\cdot \bs{h}_{\bs k} }$, shown at Fig. \ref{fig: energy-DOS-entropy}(a) along high symmetry points shows. The gap at the Dirac points is topological \cite{mcclarty2022topological} and proportional to the DM strength, while $\Omega$ determines the gap at the $\Gamma$ point. We use parameters for the 2D Van der Waals magnet CrI$_3$ \cite{lee2020fundamental,soriano2020magnetic}, $J = 0.2$ meV, $s=3/2$, lattice constant $a_0=6.95 \text{\AA}$, and we set the external magnetic field $B=0$.

The magnon gas is considered a thermodynamical system under the assumption that the equilibration length for interactions between magnons is much shorter than the system size \cite{dzyapko2017magnon,demidov2007thermalization}. Thus, the system is parametrized by a temperature $T$ and a chemical potential $\mu$. In addition, it is assumed that strong inelastic spin-conserving processes fix the temperature of magnons to the temperature of phonons. The rate of equilibration for the temperature of magnons with the phonon system is mainly dominated by magnon-conserving and magnon-nonconserving scattering processes \cite{streib2019magnon} and therefore, the magnon temperature equilibrates faster with respect to the magnon chemical potential. The internal energy for the magnon system is $ U = \int  d\epsilon\rho(\epsilon) n(\epsilon, T)$, with $n(\epsilon, T)$, the Bose-Einstein distribution and $\rho(\epsilon)$ the magnon density of states (DOS), while the entropy is
\begin{align}\label{entropybosons}
 S \nonumber= - k_{B}\int d\epsilon \rho(\epsilon)&\left[\ln \left[2\sinh\left[\frac{\epsilon-\mu}{2k_{B}T}\right]\right]\right.\\
&\quad+\left.\left(\frac{\epsilon-\mu}{2k_{B}T}\right)\coth\left[\frac{\epsilon-\mu}{2k_{B}T}\right]\right],
\end{align}
with $k_{B}$ the Boltzmann constant. The magnonic DOS is determined from the band structure and displayed in Fig. \ref{fig: energy-DOS-entropy}(b); see Supplemental Material (SM) for details. As expected, varying the magnetic anisotropy, or magnetic field, shifts the DOS towards higher energies. At finite DM strength, the DOS becomes null along the gapped region existing at Dirac points and reaches larger values as the gap widens. The entropy and internal energy as a function of temperature and DM coupling are shown in Fig. \ref{fig: energy-DOS-entropy}(c) and \ref{fig: energy-DOS-entropy}(d), respectively. At zero DM strength, both quantities are symmetric. However, the entropy is maximum along the constant entropy contours while the internal energy is minimum.

{\it Magnonic thermal cycle.--} Now, we focus on the performance of the magnon-based thermal machine. As proof of concept, we consider the Otto cycle for a working medium of 2D magnons, illustrated in Fig. \ref{fig:OttoCycle}, using an entropy-DM ($S-D$) diagram. The cycle comprises four stages. Initially (\textrm{1}$\rightarrow$ \textrm{2}), the magnon gas is prepared in a thermal state at $T = T_{C}$, being $T_C$ the temperature of the cold reservoir. Later, via an {\it isentropic expansion}, the system is disconnected from the thermal reservoir and experiences an adiabatic change of the DM coupling from $D_{0}$ to $D$. This stage finalizes with temperature $T_{\textrm{2}}$ that satisfies $S(T_{C}, D_{0}) = S(T_{\textrm{2}}, D)$ and thus, the  work is determined by $W_{\textrm{1}\rightarrow\textrm{2}}=U_{\textrm{2}}(T_{\textrm{2}}, D) - U_{\textrm{1}}(T_{\textrm{C}}, D_{0})$. In the second stage (\textrm{2}$\rightarrow$ \textrm{3}), an \textit{isochoric heating} takes place. The magnon system is connected to a hot reservoir and thermalizes at temperature $T_{\textrm{H}}$. As a result, only the heat flux $Q_{in}=U_{\textrm{3}}(T_{\textrm{H}}, D)- U_{\textrm{2}}(T_{\textrm{2}}, D)$ is involved. Thirdly (\textrm{3}$ \rightarrow$ \textrm{4}), through an \textit{isentropic compression}, the system is decoupled from the hot reservoir, and the DM strength varies \textit{isentropically} from $D$ to $D_{0}$. The process ends with a temperature $T_{\textrm{4}}$ satisfying $S'(T_{H}, D) = S'(T_{\textrm{4}}, D_0)$ and a work given by $W_{\textrm{3}\rightarrow\textrm{4}}=U_{\textrm{4}}(T_{\textrm{4}}, D_{0}) - U_{\textrm{3}}(T_{\textrm{H}}, D)$.  Finally (\textrm{4}$ \rightarrow$ \textrm{1}), via an \textit{isochoric cooling}, the magnon gas is again put in contact with a cold thermal reservoir at constant DM  $D_{0}$, with the heat flow given by $Q_{out}=U_{\textrm{1}}(T_{\textrm{C}}, D_{0}) - U_{\textrm{4}}(T_{\textrm{4}}, D_{0})$. In terms of the total work, $W_{T}=W_{\textrm{1}\rightarrow\textrm{2}}+W_{\textrm{3}\rightarrow\textrm{4}}$ and the heat flows between the reservoirs and working substance, we will next analyze the characteristic of the proposed thermal cycle.

\begin{figure}[ht]
\centering
\includegraphics[width=\columnwidth]{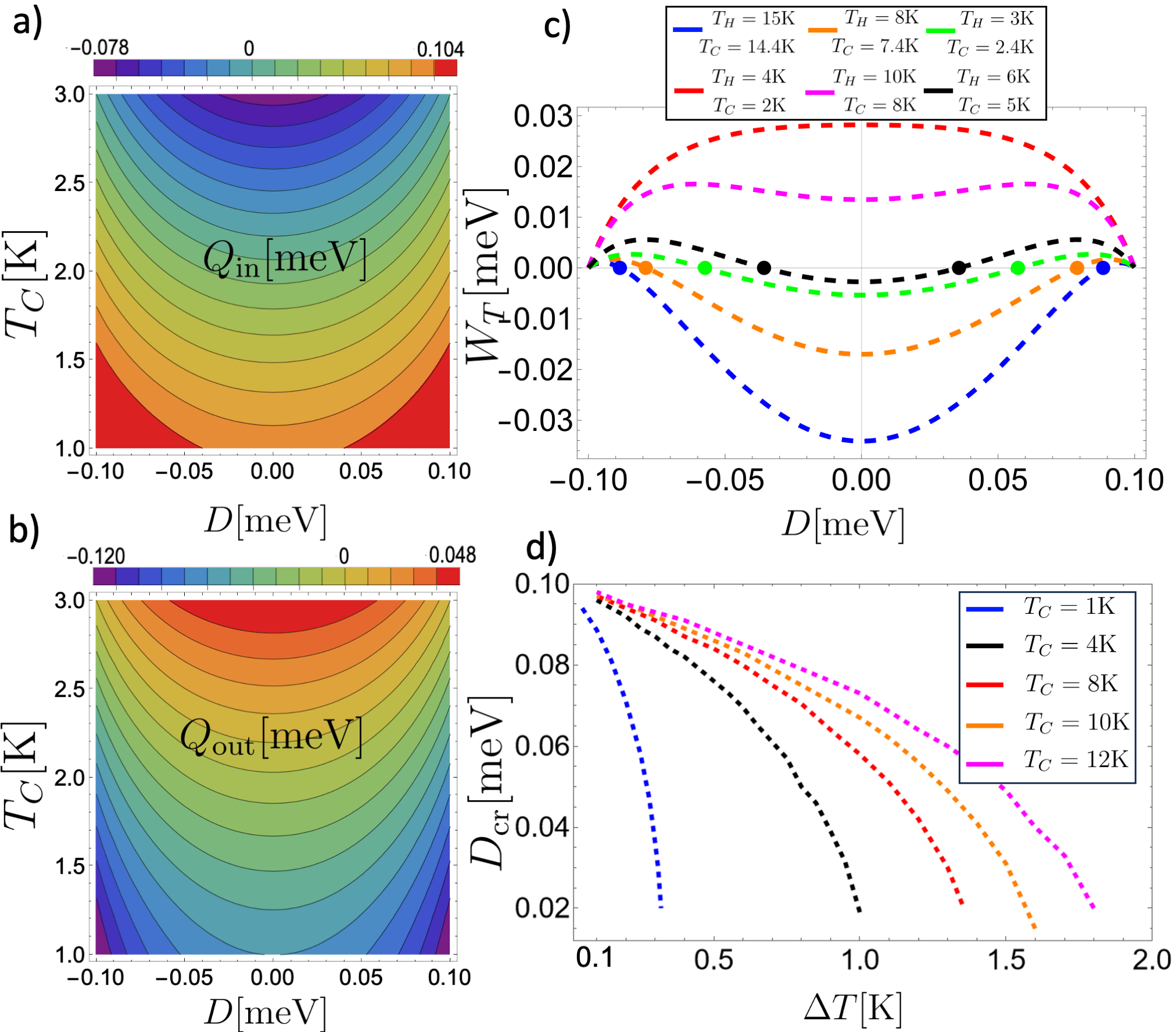}
\caption{Heat flow $Q_{in}$ and $Q_{out}$, at panels (a) and (b), respectively, as a function of DM and temperature $T_{C}$ for a fixed temperature of the hot reservoir $T_{H}=15$ K. In (c)  the total work $W_{T}$ is calculated for different $T_{H}$ and $T_{C}$. The thicker dots indicate the critical DM coupling, at which the transition between the engine and the refrigerator (where $W < 0$) behavior occurs. (d) Critical DM coupling $D_{\text{cr}}$ as a function of $\Delta T = T_H - T_C$ for different values of $T_C$.} 
\label{fig: heatflow-work}
\end{figure}

The magnonic Otto cycle exhibits two facets: an engine and a refrigerator regime, displayed in Fig \ref{fig: heatflow-work}. The engine is characterized by positive work output ($W_{T} > 0$), where the heat flows from the hot bath into the working medium ($Q_{{in}}>0$) and from the working medium into the cold bath ($Q_{{out}}< 0$). The refrigerator corresponds to negative work output ($W_{{T}} < 0$), along with heat flowing from the cold bath into the working medium ($Q_{{out}}> 0$) and from the working medium into the hot bath ($Q_{{in}}< 0$). The heat flows, $Q_{{in}}$ and $Q_{{out}}$, are shown in Fig. \ref{fig: heatflow-work}(a) and (b), respectively, while at the panel (c) the total work $W_{{T}}$ is plotted for various reservoirs temperatures. The engine and refrigerator regimes, marked by a sign change of $Q_{in}$ and $Q_{out}$, occur for a delicate combination of temperatures and DM coupling. The onset of the engine-refrigerator transition can be determined by the total work satisfying $W_{T}(D=D_{cr}, T_H, T_C)=0$, from which a relation between the critical values, $D_{cr}$, and temperature difference, $\Delta T=T_H-T_C$, is displayed at Fig. \ref{fig: heatflow-work}(d). $D_{cr}$ is the critical DM parameter for which the work is zero, except for the trivial case $D = \pm 0.1$ meV.
At low working temperatures, the transition occurs for a small difference $\Delta T$, as evidenced by the green line in Fig. \ref{fig: heatflow-work}(c), where $T_H = 3$ K and $T_C=2.4$ K, with $D_{{cr}}=\pm 0.058$ meV. 
In general, a greater temperature difference is needed for larger temperatures of the cold reservoir to fulfill the transition condition (see the black line in panel (c) as an example).  Thus, it allows us to identify quantitatively, in terms of the variables $T_C$ and $\Delta T$, the regime where the thermal cycle behaves as an engine or refrigerator.


\begin{figure}[thb]
\centering
\includegraphics[width=8.6cm]{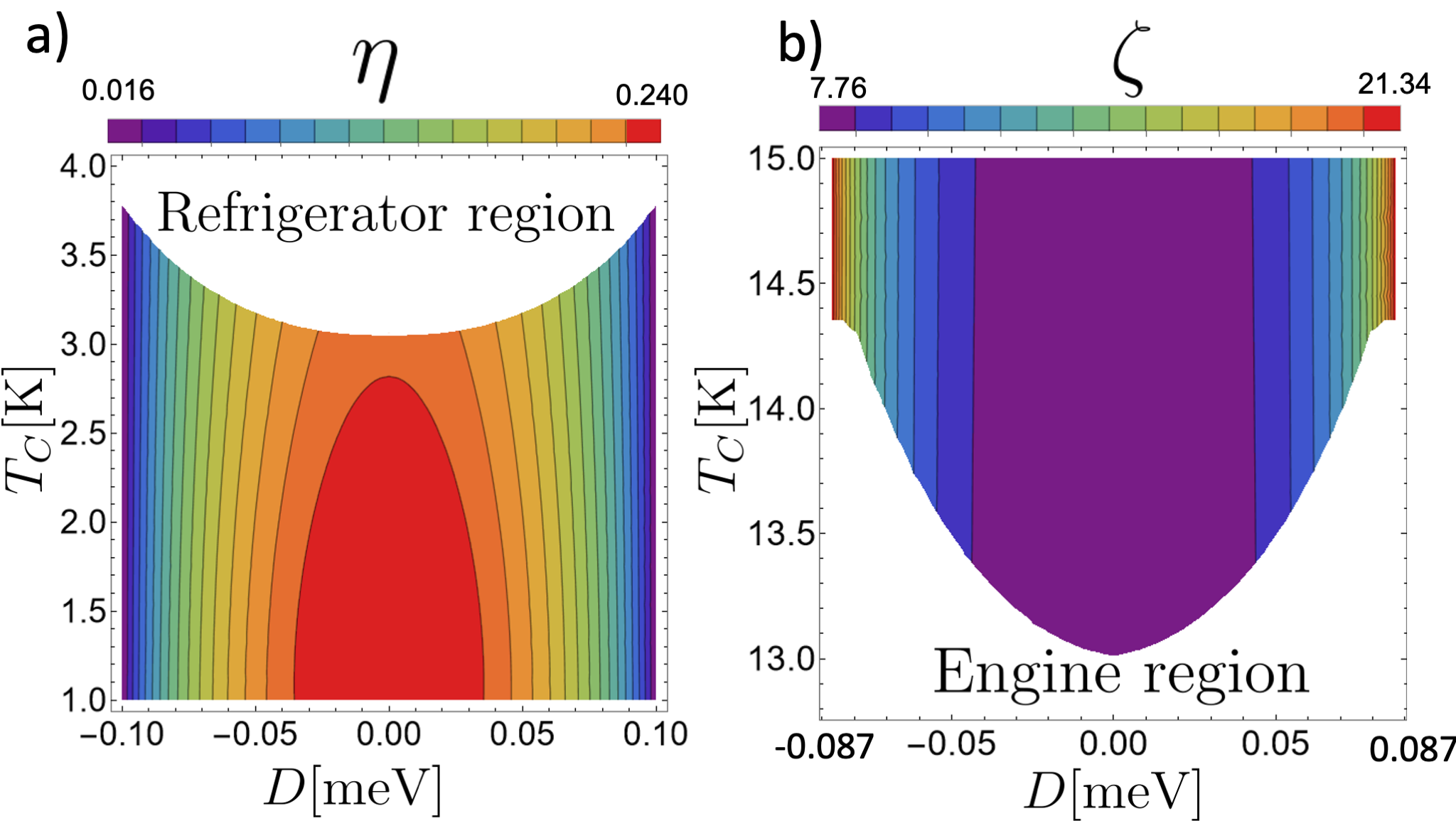}
\caption{(a) Efficiency of the magnon-based engine as a function of $D$ parameter and cold temperature of reservoir $T_{\textrm{C}}$ for a fixed $T_{\textrm{H}}=4$ K. The region where the efficiency overcomes the Carnot efficiency is called \textit{refrigerator region}, where the total work is negative. (b) Coefficient of performance $\zeta$ for the magnon-based refrigerator machine as a function of $D$ parameter and $T_{\textrm{C}}$ for a fixed $T_{\textrm{H}}=15$ K. The region blanked corresponds to the region where the total work is positive, so it is an engine region. }
\label{fig:etacop}
\end{figure}

Different coefficients capture the performance of each magnonic Otto cycle, the efficiency $\eta = \left|W_{T}/Q_{in}\right|$ for the engine, and the coefficient of performance $\zeta=\left|Q_{out}/W_{T}\right|$ for the magnonic refrigerator. The efficiency, displayed in Fig. \ref{fig:etacop}(a) as a function of $T_C$ and  DM coupling for a fixed $T_H = 4$ K, it is symmetric and maximum at $D=0$, and reaches values of about $24\%$. Importantly, there is a region where the efficiency is no longer well defined as the system overcomes the corresponding Carnot efficiency. In such a region, the work becomes negative, and the transition to a refrigerator takes place. Note that when $T_C=2$ K, the magnonic engine properly works in the full range of the DM parameter (see also red line in Fig. \ref{fig: heatflow-work}c)). 
Similarly, Fig. \ref{fig:etacop}(b) shows the coefficient of performance as a function of $T_C$ and the DM parameter for a fixed $T_H = 15$ K. Note that $\zeta$ can be computed only in the region where the system operates as a refrigerator. In this case, $\zeta$ is minimum at $D=0$  and maximizes for $D=\pm D_{\text{cr}}$. If we focus on $T_C=14.5$ K, we can see that the refrigerator regime operates at the full range of the allowed DM parameter (for comparison, see the blue line in Fig. \ref{fig: heatflow-work}(c)). However, analogously for efficiency, there is a marked region where the system becomes an engine.

{\it Discussion.-} The Otto magnon-based thermal machine exhibits a controlled change from an engine to a refrigerator. As discussed in the SM, this transition is dominated by the difference in population of thermal magnons at the stages with constant DM, i.e., \textrm{2}$ \rightarrow$ \textrm{3} (heating) and \textrm{4}$ \rightarrow$ \textrm{1} (cooling), where a linear dependence of the magnon energy with the DM has been assumed. Thus, positive work corresponds to a larger population in the thermalization process (\textrm{2}$ \rightarrow$ \textrm{3}) toward $T_H$. This excess of magnon states represents a spin accumulation, which in turn can be pumped in the form of spin currents at the interfaces of metal-magnet heterostructures. On the other hand, the refrigerator phase is related to a decrease in the thermal population when cooling the system towards $T_C$ in the stage (\textrm{4}$ \rightarrow$ \textrm{1}). In other words, the system temperature tends to decrease as long as magnons are annihilated. Interestingly, this reversion is parametrized by a critical DM parameter.

For the proposed cycle, variations on the DM parameter correspond to the exerted work over (by) the magnonic system, i.e., we recognize that $\delta W\propto\delta D$. As discussed above, the total work depends essentially on the difference in magnon population. Therefore, we parameterize $\delta W = \mu_n \delta n $, with $\mu_n$ being a constant with units of energy, and $\delta n$ stands for variations of the magnon population. Thus, the first law of thermodynamics is written as $ \delta U = T \delta S + \mu_n \delta n$ (at a constant volume). Note the similarity with an ideal gas, being $\mu_n$ playing the role of a magnon chemical potential. In other words, the notion of \textit{work} in our system is related to magnon population changes in the heating and cooling processes, which corresponds to a manifestation of a finite magnon chemical potential. Therefore, an enhancement of magnon chemical potential is predicted where a positive work is produced in the magnonic Otto cycle. For comparison, we consider a two-dimensional Dirac-type electronic system controlled by external magnetic fields where the extracted work and efficiency are comparable to those presented here \cite{pena2020quasistatic,pena2020otto,singh2021magic,pena2019magnetic,pena2015magnetostrain,myers2023multilayer}. However, interpreting useful work in electronic systems is still under discussion \cite{myers2022quantum}. Here, we presented an explicit way to functionalize the work extracted from the magnon-based Otto cycle through the possibility of pumping magnon spin currents.

In summary, we have proposed a feasible magnon-based thermal machine whose working principle relies on control over the DM coupling and reservoir temperatures. We employed the Otto cycle and found that maximum efficiency, which corresponds to the Carnot efficiency, is obtained for specific values of DM coupling. Remarkably, the presented cycle behaves as an engine or refrigerator according to the temperature difference at which the machine operates. Thermodynamical calculations show that a larger (smaller) magnon population tends to increase (decrease) the system temperature. Under the proper choice of parameters, the proposed thermal machine promotes a magnon accumulation that enhances the magnon chemical potential, which in turn could translate into magnon spin pumping at metal-magnet heterostructures when the work is positive. In order to conduct potential experiments in such systems, the adiabatic condition of varying the DM parameter without directly affecting the electronic temperature must be considered. Although we have focused on a minimal two-dimensional model for a specific bidimensional magnet, the presented framework is general and might be extended to other spin models accounting for antiferromagnets or spin-textured systems. The increasing interest in theoretical prediction and synthesis of 2D magnetic materials \cite{torelli2020high} and the experimental realization of various quantum Otto cycles \cite{rossnagel2014nanoscale,zhang2022dynamical,nettersheim2022power} might establish the experimental feasibility of our work. {Finally, we emphasize that estimating the effects of phonons within our operating temperature range is essential to get more accurate results. Nevertheless, we predict a small phonon dependence of the main results at low temperatures.}


N. V-S. thanks funding from Fondecyt Iniciacion No. 11220046. R.E.T thanks funding from Fondecyt Regular 1230747. F.J.P. acknowledges financial support from ANID Fondecyt grant no. 1240582 and “Millennium Nucleus in NanoBioPhysics” project NNBP NCN2021 \textunderscore 021. P.V. acknowledges support from ANID Fondecyt grant no. 1240582.

\bibliography{biblio-thermomagnons}

\clearpage
\onecolumngrid
\appendix
\section{Supplemental Material}
In this supplemental material we include details of calculations for the magnonic DOS, entropy, free energy, and other thermodynamical quantities. 

\subsection{Calculation of Magnon Density of States}
\begin{figure}[ht]
\centering
\includegraphics[width=15cm]{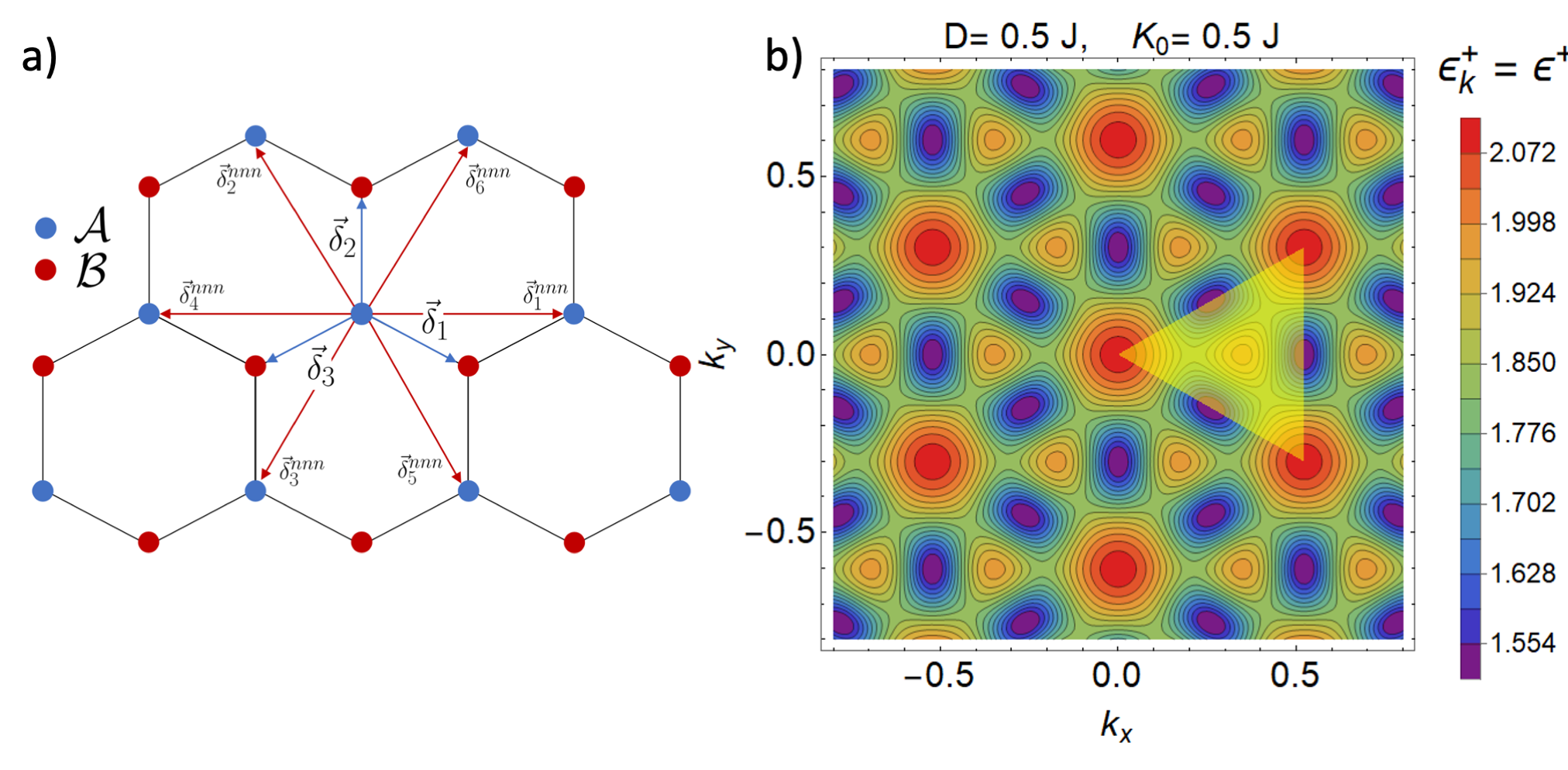}
\caption{a) Schematic of the honeycomb lattice composed of the sublattices $\mathcal{A}$ and $\mathcal{B}$ with the corresponding nearest and next-nearest neighbor vector links. b) Magnon energy $\epsilon_{\bs{k}}^+$ in the Brillouin Zone for $D = K_0= 0.5 J$ at $T=0 K$. The highlighted triangle depicts the 1BZ where the DOS is calculated. The lattice vectors are ${\bs \delta}_{1}=a_0\left(\sqrt{3}/2,-1/2\right)$, ${\bs \delta}_{2}=a_0\left(0,1\right)$, ${\bs \delta}_{3}=-a_0\left(\sqrt{3}/2,1/2\right)$}
\label{fig:dispBZ}
\end{figure}

To calculate the DOS of the system, we use a fine mesh of about $6\times 10^{6}$ $\bold{k}$-points in the area enclosed by the yellow triangle of Fig. \ref{fig:dispBZ}b), and for every $\bold{k}$-state we evaluate the energy levels coming from each band. Then, we obtain the density of states by making the energy histogram with a resolution of 15 $\mu$eV. This will give us a non-normalized density of states that we must normalize for thermodynamic calculations. The criterion we use for this goes hand in hand with the case of phonons, where the integral density of states must be equal to the number of normal modes. Since there are two magnon modes per unit cell, every normalization factor is calculated as $2/\langle N \rangle$, where $\langle N \rangle =\int _{\text{min} \,  \rho(\epsilon)}^{\text{max} \,  \rho(\epsilon)} d\epsilon \; \rho(\epsilon)$.

\subsection{Thermodynamic quantities}
Once the DOS of the system is calculated, we can use the continuum approximation to calculate the entropy of the proposed system. The general expression for the entropy of a bosonic system is given by:

\begin{equation}
\label{entropyeq2}
S = - k_{B}\int d\epsilon \; \rho(\epsilon)\left(n(\epsilon)\ln n(\epsilon) - \left(1+ n(\epsilon)\right)\ln \left(1+ n(\epsilon)\right)\right),
\end{equation}
where

\begin{equation}
\label{numbereq}
n(\epsilon,\mu,T)=\frac{1}{e^{\frac{\left(\epsilon-\mu\right)}{k_{B}T}} - 1}
\end{equation}
represents the occupation number for energy $\epsilon$ at a given temperature $T$ and chemical potential $\mu$, and $\rho(\epsilon)$ is the density of states. For this case, the magnon energy is given by $\epsilon^{\pm}(\bs k)=  \Omega \pm\sqrt{\bs {h_k}\cdot \bs {h_k}}$, with
\begin{align}
{\bs h}_{\bs k}=s
\begin{pmatrix}
-J\sum_{\eta}\cos\left[{\bs k}\cdot{{\bs\delta}_{\eta}}\right]\\
J\sum_{\eta}\sin\left[{\bs k}\cdot{{\bs \delta}_{\eta}}\right]\\
2D\sum_{\eta}\sin\left(\bs {k}\cdot\bs {\delta}_{\eta}^{nnn}\right)
\end{pmatrix},
\end{align}
and $\Omega = 3Js + 2Ks + B$. Since its number is non-conserved, we treat the magnonic system as a gas of non-interacting bosons with the chemical potential fixed at the lowest energy value in the band structure. This means we work in a formulation with constant magnon chemical potential. Next, the expression inside the integral accompanying the density of state in Eq.~(\ref{entropyeq2}) can be compacted, and the entropy takes the form:

\begin{align}
\label{entropybosons2}
S = - k_{B}\int d\epsilon \rho(\epsilon)\left(\ln \left[2\sinh\left[\frac{\epsilon-\mu}{2k_{B}T}\right]\right] 
+ \left(\frac{\epsilon-\mu}{2k_{B}T}\right)\coth\left[\frac{\epsilon-\mu}{2k_{B}T}\right]\right).
\end{align}

The expression for the internal energy of the boson system is given by:

\begin{equation}
\label{internalenergy}
    U = \int  d\epsilon \; \rho(\epsilon) n(\epsilon,\mu,T).
\end{equation}


In our thermodynamic analysis, it is important to recall that the total entropy $(S)$ for this model can be written as
\begin{equation}
\label{totalentropy}
 {S = S_{m} (T,D) + S_{l}(T)},
\end{equation}
where $S_{m}(T,D)$ is the pure magnonic entropy and $S_{l}(T)$ is the entropy of the lattice related to the contribution of phonons in the system. Eq.~(\ref{totalentropy}) assumes that the entropy of phonons relies solely on temperature, thus neglecting the influence of phonon coupling with external magnetic fields or the DM interaction. Furthermore, for the comprehensive assessment of entropy, we disregard magnon-phonon interactions and assume that the temperature regime in our system is low enough so that the general results are not affected by phononic entropy.

\subsection{Transition between engine and refrigerator}

The transition between engine and refrigerator is equivalent to exploring the change in the sign of the total work. It can be analyzed through the system's internal energy if we explicitly analyze the work in the adiabatic stages. In the isentropic expansion, we have:
\begin{equation}
    W_{\textrm{1}\rightarrow\textrm{2}}=U_{\textrm{2}}(T_{\textrm{2}}, D) - U_{\textrm{1}}(T_{\textrm{C}}, D_{0}),
\end{equation}
that can be written explicitly as

\begin{equation}
\label{workida}
      W_{\textrm{1}\rightarrow\textrm{2}}= \int_{\epsilon_{min}}^{\epsilon_{max}} d\epsilon \; \epsilon \left[\; \rho_{\textrm{2}}(\epsilon) \; n_{\textrm{2}}(\epsilon, T_{\textrm{2}}, \mu) - \rho_{\textrm{1}}(\epsilon) n_{\textrm{1}}(\epsilon, T_{C}, \mu)\right],
\end{equation}
where we have used that the range of integration in the internal energy is approximately the same at points 1 and 2. In the same spirit, the work on adiabatic compression will have the expressions 

\begin{equation}
    W_{\textrm{3}\rightarrow\textrm{4}}=U_{\textrm{4}}(T_{\textrm{4}}, D_{0}) - U_{\textrm{3}}(T_{\textrm{H}}, D),
\end{equation}
or 
\begin{equation}
\label{workcomeback}
      W_{\textrm{3}\rightarrow\textrm{4}}= \int_{\epsilon_{min}}^{\epsilon_{max}} d\epsilon \; \epsilon \left[\; \rho_{\textrm{4}}(\epsilon) \; n_{\textrm{4}}(\epsilon, T_{\textrm{4}}, \mu) - \rho_{\textrm{3}}(\epsilon) n_{\textrm{3}}(\epsilon, T_{H}, \mu)\right].
\end{equation}

If we consider the cycle in terms of the density of states, we have, according to the isochoric trajectories, the next condition: 

\begin{equation}
\label{conditionrho}
    \rho_{\textrm{1}}(\epsilon) = \rho_{\textrm{4}}(\epsilon) \; \text{;} \; \rho_{\textrm{3}}(\epsilon) = \rho_{\textrm{2}}(\epsilon).
\end{equation}
Eq.~(\ref{conditionrho}) can be used in combination with Eqs.~(\ref{workida}) and (\ref{workcomeback}) to write down the total work:
\begin{equation}
\label{wfinal}
    W_{\text{T}} = -\left[\int_{\epsilon_{min}}^{\epsilon_{max}} d\epsilon \; \epsilon \; \left\lbrace \rho_{\textrm{2}}(\epsilon) \left[ n_{\textrm{2}}(\epsilon, T_{\textrm{2}}, \mu)- n_{\textrm{3}}(\epsilon, T_{H}, \mu)\right] +  \rho_{\textrm{1}}(\epsilon) \left[ n_{\textrm{4}}(\epsilon, T_{\textrm{4}}, \mu)- n_{\textrm{1}}(\epsilon, T_{C}, \mu)\right] \right\rbrace \right],
\end{equation}
where $T_{\textrm{2}}$ and $ T_{\textrm{4}}$ are obtained from the isentropic trajectory conditions:
\begin{equation}
\label{isentropicfinal}
    S(T_{C}, D_{0}) = S(T_{\textrm{2}}, D) \; , \; S(T_{H}, D) = S(T_{\textrm{4}}, D_{0}).
\end{equation}

\begin{figure}[ht]
\includegraphics[width=0.8\textwidth]{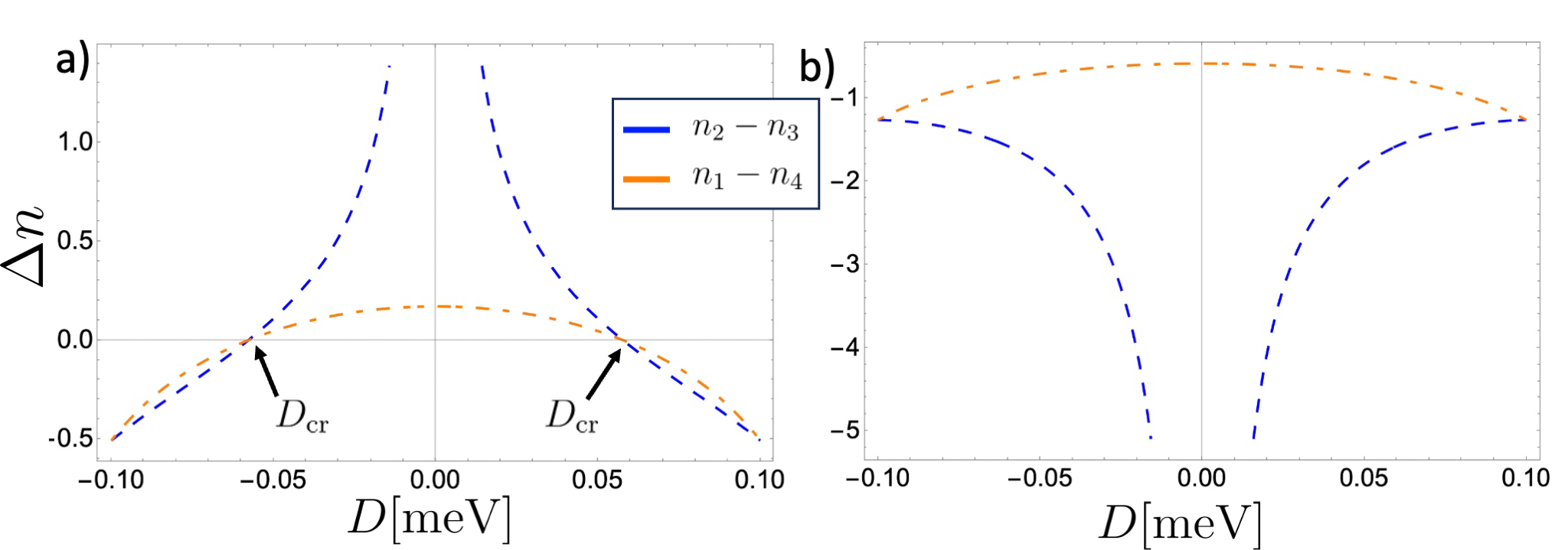}
    \caption{Graphical solution of Eq. \eqref{eq:conditionwork} for the case $\frac{\rho_{\textrm{1}}(\epsilon)}{ \rho_{\textrm{2}}(\epsilon)}=1$ as a function of the DM parameter for (a) $\Delta T= 0.6$ with $T_H = 3$ K, and (b) $\Delta T= 5$ with $T_H = 15$ K }
    \label{population}
\end{figure}

From Eq.~(\ref{wfinal}), we can find the critical relationship when a point of zero work occurs, that is, a point where a reversal in the behavior of the proposed engine will emerge. Therefore, the expression of this critical point is given by

\begin{equation}
\label{eq:conditionwork}
 n_{\textrm{2}}(\epsilon, T_{\textrm{2}}, \mu)- n_{\textrm{3}}(\epsilon, T_{H}, \mu) = - \frac{\rho_{\textrm{1}}(\epsilon)}{ \rho_{\textrm{2}}(\epsilon)} \left[ n_{\textrm{4}}(\epsilon, T_{\textrm{4}}, \mu)- n_{\textrm{1}}(\epsilon, T_{C}, \mu)\right].
\end{equation}
Once $T_{\textrm{2}}$ and $T_{\textrm{4}}$ of the isentropic trajectories are correctly parameterized via Eq.~(\ref{isentropicfinal}), and under the assumption that $\frac{\rho_{\textrm{1}}(\epsilon)}{ \rho_{\textrm{2}}(\epsilon)} \sim \text{const.}$, we can numerically explore the population differences $n_{\textrm{2}}-n_{\textrm{3}}$ and $n_{\textrm{1}}-n_{\textrm{4}}$ as a function of the parameter $D$ by considering that in real space the magnon energy follows $\epsilon\sim \mathcal{C} D$ (with $\mathcal{C}$ an arbitrary constant). In Fig. \ref{population}(a), we graphically solve Eq. \eqref{eq:conditionwork} as a function of the DM parameter for the case $\frac{\rho_{\textrm{1}}(\epsilon)}{ \rho_{\textrm{2}}(\epsilon)}=1$. Valid solutions correspond to a crossing between $n_{\textrm{2}}-n_{\textrm{3}}$ and $n_{\textrm{1}}-n_{\textrm{4}}$, excepting the trivial case at $D = \pm 0.1$ meV, where the cycle starts (finishes). Such a crossing occurs exactly at $D = \pm D_{\text{cr}}$, where a transition from the engine to the refrigerator behavior takes place for the parameters $T_{H}= 3$ K, $T_{C}=2.4$ K, i.e., $\Delta T=0.6$, which effectively gives rise to a change in the behavior of the thermal machine, as mentioned in the main text. In contrast, from Fig. \ref{population}(b), we do not observe any cross on the difference of thermal populations (so that there is not a valid solution of Eq. \eqref{eq:conditionwork}) for the temperatures $T_{H}= 15$ K and $T_{C}=10$ K, i.e., $\Delta T=5$. In this case, since the enclosed area by $n_{\textrm{2}}-n_{\textrm{3}}$ it is always bigger than the corresponding one by $n_{\textrm{1}}-n_{\textrm{4}}$ (in absolute value), we have that $W_{\text{T}} > 0$ and the machine operates as an engine.

\subsection{Calculation of critical $T_C$ for the transition between engine and refrigerator at $D=0$}
\begin{figure}[h!]
\centering
\includegraphics[width=8cm]{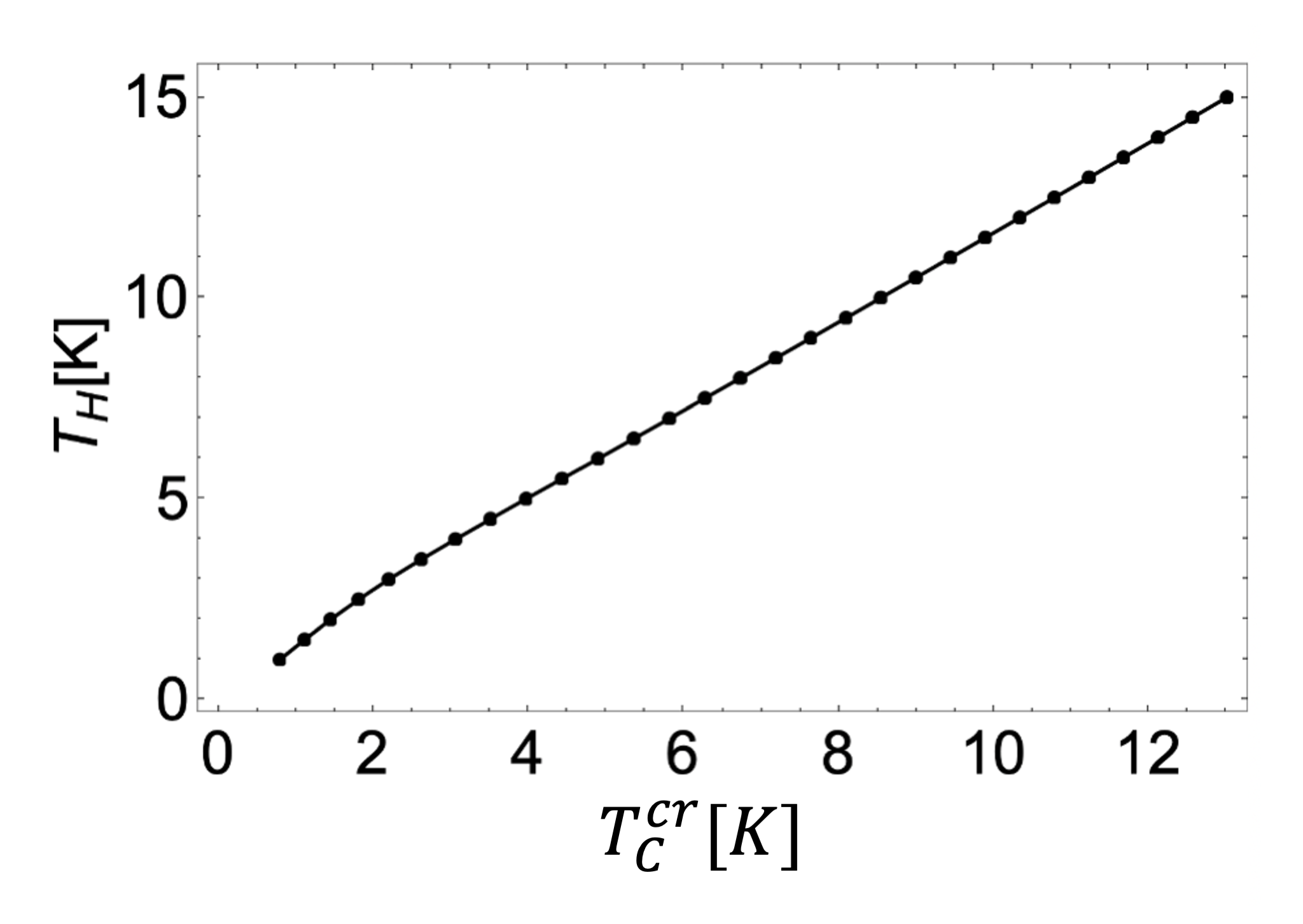}
\caption{High temperature of the reservoir as a function of the minimum cold temperature of the reservoir so that the transition between engine and refrigerator occurs.}
\label{fig:TL_critical}
\end{figure}

Here, we show the existence of a critical $T_L$ so that, given a certain $T_H$, the transition between an engine and refrigerator takes place. We first find the minimum DM parameter at which the total work becomes zero. As shown in Fig. 3 in the main text, the total work curves that admit the transition between an engine and refrigerator always have a global minimum at $D=0$. Therefore, at $D=0$, one can ensure that a transition can occur for some set of parameters (see also Fig. 4 of the main text). Next, we explore the temperatures needed to achieve such a transition. To do that, we fix a given $T_H$ and calculate what is the minimum $T_C$ to accomplish the condition $W_{\text{Total}}=0$. In Fig. \ref{fig:TL_critical}a), we show the relationship between different values of $T_H$ and its corresponding $T_C^{\text{cr}}$, defined as the minimum value of $T_C$ so that the transition occurs. As can be seen, the behavior is almost linear. However, from Fig. \ref{fig:TL_critical}b), where we show the \text{slope} of the curve presented in Fig. \ref{fig:TL_critical}a), it can be noticed that when $T_H$ is larger than 5 K, the behavior between $T_H$ and $T_C^{\text{cr}}$ is almost linear. At the same time, for low $T_H$, the relationship has a more complex behavior. 

\end{document}